# Research Directions in Democratizing Innovation through Design Automation, 'One-Click' Manufacturing Services and Intelligent Machines

Binil Starly, Atin Angrish and Paul Cohen
Edward P. Fitts Department of Industrial and Systems Engineering,
North Carolina State University, Raleigh, NC, USA

**Abstract**
The digitalization of manufacturing has created opportunities for consumers to customize products that fit their individualized needs which in turn would drive demand for manufacturing services. However, this 'pull' based manufacturing system production of extremely low quantity and limitless variety for products is expensive to implement. New emerging technology in design automation driven by data-driven computational design, manufacturing-as-a-service marketplaces and digitally enabled micro-factories holds promise towards democratization of innovation. In this paper, scientific, technology and infrastructure challenges are identified and if solved, the impact of these emerging technologies on product innovation and future factory organization is discussed.

-----------------------------------------------------------------------------------------------------------------

## 1. Introduction

The ability to democratize innovation across users with a wide range of skills and experience is enabled by the convergence of advances made in cyber-information engineering, manufacturing process technologies and the social science domains. Democratizing innovation means that tools for users and consumers to engage in product design for customization are available and accessible [1]. Democratizing innovation can also lead to entirely new paradigms of expanding the typical profile of a manufacturer to also include those who operate micro-factories, leading to the prospect of having customized products built anywhere and anytime.

Computing technology has created orders of magnitude efficiency in the product life cycle but the skills required to design products have been largely confined to those skilled in the art and science of design and making of things. If barriers to lowering skills needed to engage in product design are reduced, an increased expansion of the innovation ability of the consumer base will emerge [2-5]. Products can be designed by anyone and not necessarily limited to those skilled in engineering and industrial design. Lowering the barriers would also mean that humans can focus on innate creativity, while computing tools work behind the scenes to improve designs and make recommendations while working collaboratively with the human. Once the product is designed and verified, consumers can potentially engage in a 'one-click' interaction to have the product assembly manufactured and delivered back to the client.

An example scenario is depicted in Figure 1. Other examples can include spare parts, personalized medical devices, customized consumer products, etc. Assistive design tools can enable a child to simply sketch out or verbally describe an imaginative toy while algorithms automatically detect design intent and generate a 3D Product model (a full Boundary representation - B-rep model) of the child's version of the toy. Lowering the barriers would also allow a parent to engage in 'one-click' interaction to have the toy built. Algorithms would automatically determine the best capable and available manufacturers to make the product, including linking various manufacturers to suggest a price and delivery date. Machines negotiate and accept job order requests, fulfilling order and initiating last-mile logistics [6-8] to ship product to the client. With computing algorithms and physical machines performing the bulk of the operational steps, the focus of human activity will shift away from the requisite knowledge of design and manufacturing tools but towards innate human creativity in product design.

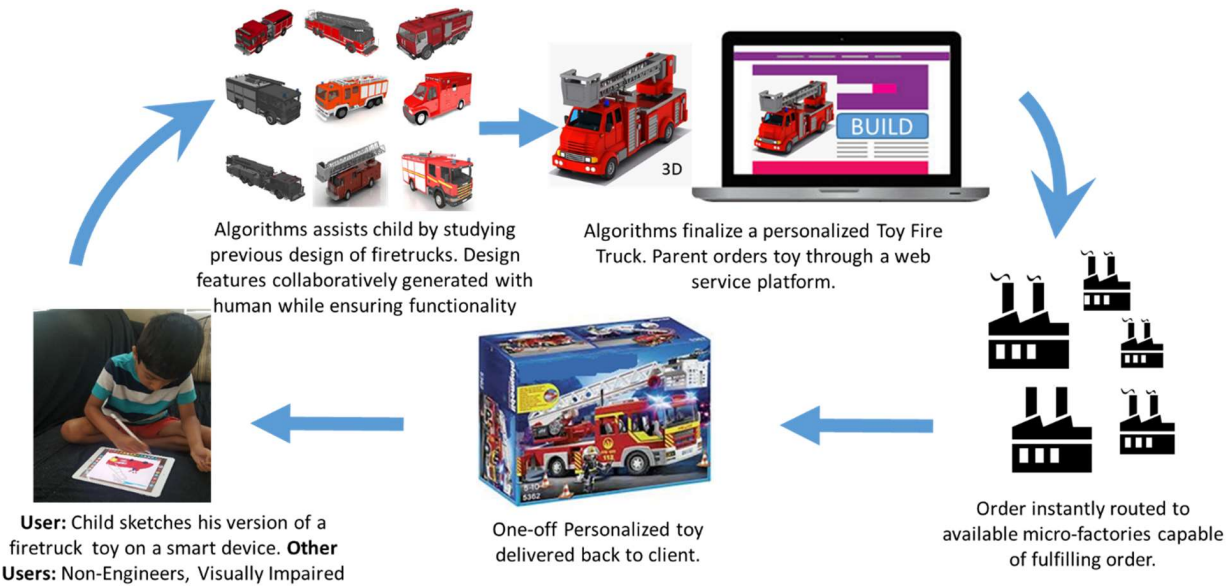

*Figure 1: Design by Anyone, Build Anywhere, Anytime, for Highly Customizable, One-Off, and Complex Products*

## 2. Enabling Technologies for Democratizing Design and Manufacturing

This section describes three areas in which advancements must be made to democratize design and manufacturing of heavily customized products (summarized in Table 1).

### 2.1. Collaborative Design Bots for Product Design

For computing systems to augment human creativity in product design, several engineering hurdles must be overcome. Algorithms must take in high-level human input, such as functional requirements of a product, and then be able to synthesize design intent to create a first draft of a featured digital product model. Combined with human input, design bots could iterate through hundreds of design options with humans driving the product design. To adequately advance the knowledge base to build 'Collaborative Design Bots', advancements must be made in the field of Natural Language Processing (NLP) and Computational Linguistics with design and manufacturing based vocabulary. Recent developments in neural machine translation (NMT) are already approaching human level comprehension [9-11] for understanding text and parsing information to power collaborative design bots in a Text-2-CAD paradigm (Figure 2).

New advancements must be made in training computers to generate complex graphs as those seen in the internal mathematical representations of Engineering CAD models [12-13]. By designing algorithms to learn graphs, automation can be built into the design process for routine tasks thereby significantly aiding productivity of users. Therefore, if algorithms that can learn product design features by ingesting millions of existing product designs to generate new design recommendations are made, they can play the digital enabler for individuals' uninitiated in product manufacturing. Concurrently, to improve the accessibility of design software, augmented reality and haptic tools [14-19] must be developed to allow non-conventional users, such as visually impaired to 'feel' 3D shapes. If computing algorithms automate mundane tasks, research must be performed to find ways in which humans can train to improve their own creativity required to design new products. For serious Artificial Intelligence (AI) in digital product design, reproducing entirely new design concepts never seen before would fundamentally change the human-technology frontier in terms of the product design process.

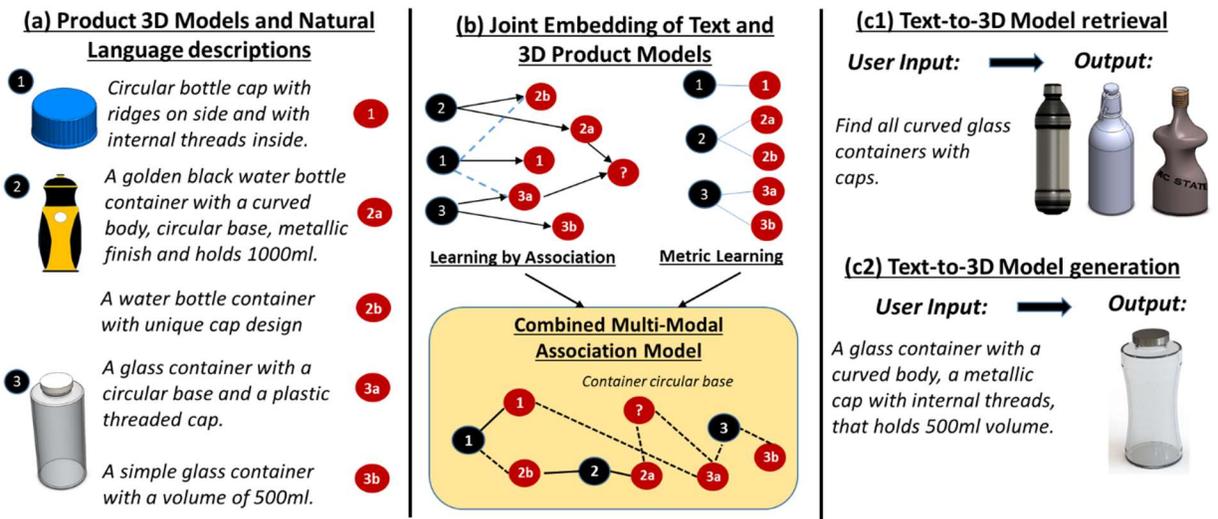

*Figure 2: Text-2-CAD paradigm (a) If datasets that leverage pairing of natural language descriptions of products are made available; (b) a joint embedding of textual description and associated 3D models can be made which clusters similar 3D shapes and text together through semantic relationships, engineering specifications and/or shape descriptors; With a jointly embedded model, two application arise - (c1) Users can search for 3D product models based on detailed text descriptions of the product model. (c2) Design bots can be able to synthesize new designs based on textual descriptions provided by the user.*

## 2.2. "One-Click" Manufacturing Service Marketplaces

While several manufacturing-as-a-service marketplaces [20-22] have operated recently, these platforms are centralized intermediary brokers responsible for allocating orders to independent service providers. To realize the scenario of a "One-Click manufacturing" marketplace, several fundamental advancements must be made in the design of decentralized two-sided marketplaces that connects users with capable service providers. Manufacturing data remains the closely guarded asset that links the two sides, but its value can only be leveraged when such data is shared. Therefore, new scientific and technology advancements must be made in: 1) Computing on encrypted data [23-24], such as 3D model data, machine information, pricing data etc., so that each party may share such information without fear of it being directly misused by either party; 2) Incentivization mechanisms [25-26] through micro-economic approaches and multi-class stochastic matching models to achieve fairness in the decentralized marketplace to ensure sustainable long-term operation; 3) New techniques which move computation to the end point, rather than move data to a centralized system, such as through Federated Learning [27-28], would minimize issues surrounding privacy and security while still facilitating data-driven computational algorithms; 4) Legal strategies and policy approaches to ensure that digital manufacturing marketplaces can thrive in an inter-connected world. Eventually "*One-Click*" Manufacturing intends to build a truly decentralized system that aims to reduce uncertainty in manufacturing networks by closing the gap between design users and the industrial base characterized by micro-factories, small job shops, and related manufacturing service providers.

*Table 1: Attributes of enabling technologies for democratizing design and manufacturing with their rationale and selected research needs to achieve that vision.*

| | **Attribute** | **Rationale** | **Research Needs for Realization** |
|---|---|---|---|
| Collaborative Design Bots for Product Design | Augment human creativity. | Engage engineers and non-engineers; Empower entrepreneurs and small businesses. | Develop algorithms to understand high level human input. |
| | Enable new modes of computer interaction to design products. | Increase accessibility of software interfaces. | Develop and study the use of augmented reality and haptic tools. |
| | Able to search existing designs to reuse designs. | Enable efficient search for similar products. | Search combining engineering specifications and geometry/topology data; Learn preferences to suggest model families. |
| | Able to offer design changes for manufacturability. | Develop cost effective, high quality, manufacturable products. | Algorithms to learn from existing products; Learn the relationships between design specification and manufacturability. |
| "One-Click" Manufacturing Service Marketplaces | Effective two-sided platforms | Must have well designed platform with effective user interface. | Design of two-sided marketplaces and their user interfaces. |
| | Digital Trust and Security of Data | Computation of digital trust enables rapid formation and dissolution of supply chain partners. | Microeconomic incentivization mechanisms; Federated learning approaches for endpoint computation on encrypted data. |
| | Smart Paperless Contracts | Legal strategies are needed to develop enforceable, fair smart contracts. | Design of Smart contract structures in the context of decentralized systems such as Blockchain systems. |
| Software Defined Manufacturing Machines | Separation of software and physical components | Allows resources to be digitally controlled and agile. | New approaches to machine and controller design. |
| | Composable process plans | Allows manufacture to be planned regardless of machine employed without manual intervention. | Understanding relationship between geometry and processing. |
| | Virtualization of Machines | Allows for better planning and self-awareness | Digital twin development; Sensor-based control. |

### 2.3. Software Defined Manufacturing Machines and Enhanced Portability

To distribute manufacturing capability through automated mini-factories and micro-factories, technological advances must be made in the design architecture of physical machines (3D printers, CNC machines, laser cutters, robots and material handling systems). The separation of the software components and physical machine through a control layer is critical to make manufacturing resources digitally programmable and agile to allow dynamic adjustment of order traffic to meet changing needs through the development of Software Defined Manufacturing (SDM) technologies [29-31]. Process plans for the designs must be composable, transportable from one machine to the other with little or no manual intervention [32] similar to how programs written in high level languages such as Java or C may be run on any operating system. If automation combined with interoperable open standards is achieved, it would allow factories to be economically viable for small companies and in rural regions by cutting down on non-productive tasks such as manual supplier sourcing, assessment, certification and compliance. Advances in virtualization of machines can enable intermediary systems to shuttle design order requests to capable and available machines anywhere in the country. Limited human involvement in connecting users with machines and enhanced portability of software systems between machines that process those orders would significantly aid in cost-reduction and democratizing activity among a broader use base.

### 3. Societal Impact

Engineering solutions that enable expanding the base of those involved in design, fabrication and manufacturing services can assist to reduce economic disparity and provide an opportunity of innovation irrespective of skillset and geographic location. It will bring design, manufacturing and innovation to small companies, new entrepreneurs and even students who may lack the skills needed to achieve detailed product design and manufacturing process plans. Design automation interfaces will radically change current interfaces for CAD/CAM software and this will open up the use of product design software to be more inclusive by expanding its use to a section of the population that never had access, such as the visually or physically challenged, who are still cognitively capable of engaging in product design. To build anywhere, and anytime, means manufacturers are digitally accessible, instantly verified, their service capacity is predicted in real-time, micro-factories are digitally networked and product process plans can be simulated and optimized through virtualization of services. Network capacity can be efficiently utilized while connected marketplaces can ensure marketing costs for small manufacturers are reduced, thereby making locally sourced, highly personalized manufacturing more competitive.

### 4. Conclusions

This paper presents scientific and technology challenges with critical research directions towards needed towards democratizing design and manufacturing. The impact of implementing tools and technologies for broad use among users, consumers and factory owners requires transdisciplinary collaboration across various domains of cyberinformation, design and manufacturing process automation, economics, business, law and public policy. It has the potential to impact rural economies across the world by directly connecting consumers with resource capable factories. The developed tools can enable new growth products and markets through mass entrepreneurship.

**Acknowledgement**
The authors would like to thank the support of NSF EEC #1840363 towards identifying new game-changing research directions in cybermanufacturing.

Califorina Berkeley, June 2-3$^{rd}$ 2016.
https://www.icsi.berkeley.edu/icsi/sites/default/files/Cyber%20Manufacturing%20WorkshopReport%20-%20Final%202017-10-30.pdf